\documentclass[journal, twoside]{IEEEtran}

\usepackage[utf8]{inputenc}
\usepackage[T1]{fontenc}
\usepackage{url}
\usepackage{ifthen}
\usepackage{cite}
\usepackage[cmex10]{amsmath}
\usepackage{amssymb,amsthm}
\usepackage{dsfont}
\usepackage{mathtools}
\usepackage{amsmath}
\usepackage{xfrac}
\usepackage{stmaryrd}
\usepackage{subfigure}
\usepackage{enumerate}

\usepackage{epsfig}
\usepackage{epstopdf}

\usepackage{tikz}
\usepackage{pgfplots}

\usetikzlibrary{arrows,shapes,backgrounds,plotmarks,positioning}

\DeclareMathOperator{\argmin}{arg\,min}
\DeclareMathOperator*{\argmax}{arg\,max}

\newtheorem{theorem}{Theorem}

\newtheorem{definition}{Definition}
\newtheorem{proposition}[theorem]{Proposition}

\newtheorem{remark}{Remark}

\newtheorem{corollary}[theorem]{Corollary}
\newtheorem{example}{Example}

\newcommand{\range}[1]{\llbracket #1 \rrbracket}

\newcommand{\Leak}{\mathcal{L}}
\newcommand{\LeakS}{\mathcal{L}_\star}

\newcommand{\Set }[1]{\{#1\}}

\begin{document}
\title{Measuring Information Leakage in \\ Non-stochastic Brute-Force Guessing}

\author{Ni Ding,~\IEEEmembership{Member,~IEEE, }Farhad Farokhi,~\IEEEmembership{Senior Member,~IEEE}
\thanks{An earlier version of the results were presented in 2020 IEEE Information Theory Workshop (ITW 2020), which was held virtually in 2021~\cite{farokhi2020ITW}.}
\thanks{N. Ding is with School of Computing and Information Systems, the University of Melbourne. The work of Ni Ding is funded by the Doreen Thomas Postdoctoral Fellowship at the University of Melbourne.}
\thanks{F. Farokhi is with the Department of Electrical and Electronic Engineering, the University of Melbourne. When working on this paper, F. Farokhi was also affiliated with CSIRO's Data61. The work of F. Farokhi is, in part, funded by the Faculty of Engineering and Information Technology at the University of Melbourne.}
\thanks{emails:\{ni.ding,farhad.farokhi\}@unimelb.edu.au}
}

\maketitle

\begin{abstract} This paper proposes an operational measure of non-stochastic information leakage to formalize privacy against a brute-force guessing adversary.
The information is measured by non-probabilistic uncertainty of uncertain variables, the non-stochastic counterparts of random variables.
For $X$ that is related to released data $Y$, the non-stochastic brute-force leakage is measured by the complexity of exhaustively checking all the possibilities of the private attribute $U$ of $X$ by an adversary. The complexity refers to the number of trials to successfully guess $U$.
Maximizing this leakage over all possible private attributes $U$ gives rise to the maximal (i.e., worst-case) non-stochastic brute-force guessing leakage. This is proved to be fully determined by the minimal non-stochastic uncertainty of $X$ given $Y$, which also determines the worst-case attribute $U$ indicating the highest privacy risk if $Y$ is disclosed.
The maximal non-stochastic brute-force guessing leakage is shown to be proportional to the non-stochastic identifiability of $X$ given $Y$ and upper bounds the existing maximin information. The latter quantifies the information leakage when an adversary must perfectly guess $U$ in one-shot via $Y$.
%
%
Experiments are used to demonstrate the tradeoff between the maximal non-stochastic brute-force guessing leakage and the data utility (measured by the maximum quantization error) and to illustrate the relationship between maximin information and stochastic one-shot maximal leakage.
\end{abstract}

\section{Introduction}

There is no doubt that data privacy and security is more important than ever.
With increasing frequency and volume of data sharing activities, people (e.g., individuals, businesses, and government entities) are becoming more concerned about confidentiality of personal information and whether it can be inferred and maliciously used for discrimination or unfair decision making.
Cyber-security also faces new threats stemming from advances in data mining, machine learning, and big data analytic. These threats are exacerbated by seemingly unlimited computational resources and capabilities of adversaries, multi-party side information attacks, unpredictable uses of large collections of data, and  non-intuitive private information leakage~\cite{PrivBigData2016Shui,PrivBigData2013Meiko}.
To address these challenges, there is an urgent need to better understand privacy risk of data disclosure measured by private information leakage.

Although it is hard to reach an unanimous agreement on the definition of data privacy and leakage~\cite{PrivBigData2016Shui,PrivMeasure2018}, fundamental requirements of an operational measure of information leakage were listed in \cite{issa2018operational}; see Appendix~\ref{app:axiom} for the exact axioms. The requirements not only point out that (R1) the measure should objectively quantify difficulty level of inferring private information but also highlights requiring minimum assumptions (R2) and accordance with intuition (R4). We believe that these requirements have not been fully addressed previously and should be revisited in the face of new challenges.

Data regulation, processing, or sanitation schemes must be applied before releasing private data and so the privacy metric is usually defined under some generic assumptions about the adversary, e.g., an adversarial model that describes the attacker's computational capability, targeted sensitive attribute, and side information~\cite{PrivMeasure2018}.
However, in complex cross-sectional multi-purpose data-sharing environments, it is hard to predict the adversary's behavior in advance. Therefore, (R2) suggests minimizing the assumption about the adversary so that the privacy measure can be generally applied.
In this sense, instead of one-shot malicious estimation as assumed in most existing studies, e.g., \cite{issa2018operational,liao2018tunable},
we propose considering an adversary with unlimited computational power, who is capable of exhaustively guessing the private data and testing it. This motivates the brute-force guessing framework of this paper.

Another existing assumption is that the inference is statistical and the side information refers to statistical prior belief. However, privacy leakage can also be non-probabilistic.
For instance, in algorithmic information theory,
the adversary's uncertainty, i.e., the difficulty in protecting a sensitive attribute, can be described by features independent of probability distribution, e.g., the size of the alphabet and the concurrence of two distinct values.
In \cite{ding2019developing}, it is elaborated that a non-stochastic framework is more desirable for small datasets using which we can only empirically determine the presence of a variable instead of its relative frequency. In other cases, particularly adversarial scenarios, uncertainty may not be described by \textit{a priori} known stochastic assumptions and must be treated as
bounded unknown variables without statistical assumptions~\cite{linsenmayer2017stabilization,teixeira2012attack}. Also, un-truthfulness in randomized privacy-preserving reports~\cite{bild2018safepub,Poulis2015} and complications in financial auditing and fraud detection~\cite{bhaskar2011noiseless, nabar2006towards} can motivate non-stochastic privacy-preserving techniques. These motivate the need to investigate information leakage in non-stochastic frameworks and to develop noiseless policies for privacy preservation by minimizing non-stochastic information leakage.

Noting catastrophic consequences of high-risk privacy breaches, a privacy measure should not undervalue the severity of security and privacy breaches, as stated in (R4), to guarantee the strength of protection schemes.
Therefore, worst-case infringements should be identified.
For example, the differential privacy \cite{Dwork2011DP} identifies the largest statistical distance between two individuals' data, i.e., the easiest pair that can be distinguished in the disclosed data set. The measures proposed in \cite{liao2018tunable,li2018privacy} quantify the worst-case information leakage when an adversary tries to infer any sensitive parameter from the disclosed data. In this paper, we consider the worst-case adversarial behaviours from the perspectives of brute-force attacks (unlimited computational capabilities) and non-stochastic information leakage (not needing statistical assumptions).

\textbf{Contributions}: In this paper, we propose a measure of information leakage in a non-stochastic guessing framework.
Assume that $X$ is the data to be released to the public after privacy-aware sanitation or processing. The processed data is  $Y$, which will be eventually released to the public. We consider a brute-force guessing setup in which an adversary is assumed to have unlimited computational resources that can exhaustively guessing the sensitive attribute $U$ of $X$, i.e., checking all the possibilities of $U$ that are compatible with its observations to find out the actual realization, by access to $Y$.
This is similar to the interpretation of~\cite{issa2018operational} for password guessing or side-channel attacks on cipher systems in which an adversary can repeatedly check all the possible combinations in the disclosed data.
But, the inference is assumed to be non-probabilistic, for which we use uncertain variables, the non-stochastic counterparts of random variables introduced in~\cite{nair2013nonstochastic}, to quantify the information leakage.

1) Considering the case when the adversary is inferring an attribute $U$ of $X$, we propose the non-stochastic brute-force guessing leakage as the ratio of the worst-case number of guesses for the adversary in the presence of the output $Y$ and in the absence of it. This definition is consistent with the stochastic brute-force guessing leakage~\cite{osia2019privacy} with the exception of {avoiding} distributions or statistics.

2) Relaxing the assumption that the adversary targets a specific attribute $U$, i.e., considering when we are not aware of the adversary's intentions, we propose the maximal leakage as the worst-case (i.e., the largest) non-stochastic brute-force guessing leakage over all attributes $U$. %
The exact value of this maximal leakage is shown to be fully determined by the conditional uncertainty of $X$ given $Y$.
The derived maximal leakage measure satisfies not only the requirements (R1), (R2) and (R4), but also axiomatic properties in (R3): post-processing inequality, independence property, and additivity; see Appendix~\ref{app:axiom} for these axioms.
We demonstrate by an example that, for given $X$ and $Y$, the worst-case sensitive attribute $U$ that incurs the maximal non-stochastic brute-force guessing leakage is determined by the minimal uncertainty of $X$ conditioned on $Y$, defined by conditional ranges of uncertain variables~\cite{nair2013nonstochastic}.

3) We explore the relationship between the maximal non-stochastic brute-force guessing leakage and the existing measures of information leakage in the literature. First, maximal non-stochastic leakage is proportional to the non-stochastic identifiability of $X$ via the observation of $Y$. Second, the maximin information~\cite{nair2013nonstochastic} is shown to measure the worst-case brute-force guessing leakage over all $U$ such that the adversary must always correctly guess the attribute in the first guess via the observation $Y$. Therefore, the maximin information is upper bounded by the maximal non-stochastic brute-force guessing leakage. Recently, it was shown that the maximin information captures the entropy of common uncertain variable (the largest uncertain variable that can be directly computed using $X$ and $Y$) and is thus useful for understanding private information in perfect privacy~\cite{9457645}. Third, we perform experiments on a real-world dataset to observe the privacy-utility tradeoff, where it is shown that the maximal stochastic (one-shot) guessing leakage in~\cite{issa2018operational} is no greater than the maximin information, which is further upper bounded by the maximal non-stochastic brute-force guessing leakage.

\textbf{Organization}:
The rest of the paper is organized as follows. We present preliminary material on uncertain variables and non-stochastic information theory in Section~\ref{sec:uv}.
In Section~\ref{sec:info_leakage}, we measure information leakage from a specific sensitive attribute $U$ to the output $Y$ and use it as the privacy measure when we are aware of adversary's intentions.
We extend this notion to when we are not aware of the adversary's intentions by defining and computing maximal non-stochastic brute-force guessing leakage, the relationship of which to the non-stochastic identifiability is studied in Section~\ref{sec:privacy}.
We compare the brute-force notion of non-stochastic information leakage with one-shot guessing measures, such as maximin information and stochastic maximal leakage in Section~\ref{sec:relationship}. Finally, we show experimental results on a heart disease dataset in Section~\ref{sec:experiment} and conclude the paper in Section~\ref{sec:conc}.

\section{Uncertain Variables} \label{sec:uv}
We borrow the following definitions of non-stochastic information in~\cite{nair2013nonstochastic}. Consider uncertainty set $\Omega$. An uncertain variable, uv in short, is a mapping on $\Omega$. For example, for uv $X:\Omega\rightarrow\mathbb{X}$, $X(\omega)$ is the realization of uv $X$ corresponding to uncertainty $\omega\in \Omega$. For any two uvs $X$
and $Y$, the set $\range{X,Y}:=\{(X(\omega), Y(\omega)):
\omega\in\Omega \}\subseteq \range{X}\times \range{Y}$ is their \textit{joint range}. For uv $X$, $\range{X}:=\{X(\omega):\omega\in\Omega \}$ denotes its \textit{marginal range}.
The \textit{conditional range} of uv $X$, conditioned on realizations of uv $Y$ belonging to the set $\mathcal{Y}$, is $\range{X|Y(\omega)\in \mathcal{Y}}:=\{X(\omega):\exists \omega\in\Omega  \mbox{ such that } Y(\omega)\in\mathcal{Y}\} \subseteq \range{X}.$ If $\mathcal{Y}=\{y\}$ is a singleton,  $\range{X|Y(\omega)\in \{y\}}=\range{X|Y(\omega)\in \mathcal{Y}}$ is replaced with $\range{X|Y(\omega)=y}$ or $\range{X|y}$ when it is clear from the context. For any two uvs $X$ and $Y$, we define the notation  $\range{Y|X}:=\{\range{Y|X(\omega)=x},\forall x\in\range{X}\}$. We sometimes refer to $\range{Y|X}$ as a {non-stochastic} channel as $\range{Y|X}$ fully {characterizes the non-stochastic} communication channel from $X$ to $Y$. In this paper, we only deal with \textit{discrete} uvs possessing finite\footnote{Extension to countably infinite sets is straightforward with extra care when manipulating extended real numbers (i.e., infinity).} ranges.

Uvs $X_1$ and $X_2$ are unrelated if $\range{X_1|X_2(\omega)=x_2}=\range{X_1}$ for all $x_2\in\range{X_2}$ and \textit{vice versa}.  Similarly, $X_1$ and $X_2$ are conditionally unrelated given $Y$ if $\range{X_1|X_2(\omega)=x_2,Y(\omega)=y}=\range{X_1|Y(\omega)=y}$ for all $ (x_2,y)\in\range{X_2,Y}$. Uvs $X_i$, $i=1,\dots,n$, are \emph{unrelated} if $\range{X_1,\dots,X_n}= \range{X_1}\times \cdots \times \range{X_n}$ and \emph{conditionally unrelated} given $Y$ if $\range{X_1,\dots,X_n|Y(\omega)=y} =\range{X_1|Y(\omega)=y}\times \cdots \times \range{X_n|Y(\omega)=y}$ for all $y\in\range{Y}$. Uvs $X$, $Y$, and $Z$ form a Markov (uncertainty) chain, denoted by $X-Y-{Z}$, if $X$ and $Z$ are unrelated conditioned on $Y$, that is, $\range{X|Z(\omega)=z,Y(\omega)=y} =\range{X|Y(\omega)=y}$ for all $(z,y)\in\range{Z,Y}.$ Note that, by symmetry of the definition of unrelated uvs, $X-Y-Z$ forms a Markov chain if and only if $Z-Y-X$ forms a Markov chain. We say $X_1-X_2-\cdots-X_n$ forms a Markov chain if $X_i-X_j-X_\ell$ forms a Markov chain for any $1\leq i {<} j {<} \ell\leq n$.

Non-stochastic entropy of uncertain variable $X$ is defined as $H_0(X):=\log(|\range{X}|).$ This is often described as the Hartley entropy~\cite{hartley1928transmission, nair2013nonstochastic}, which coincides with the R\'{e}nyi entropy of order $0$ for discrete variables~\cite{sason2017arimoto,renyi1961measures}.
Conditional (or relative) entropy of uv $X$ given $Y$ is given by
$H_0(X|Y):=\max_{y\in\range{Y}}\log(|\range{X|Y(\omega)=y}|)$. This is the Arimoto-R\'{e}nyi conditional entropy of order $0$~\cite{sason2017arimoto,arimoto}. Based on this, we can define $I_0(X;Y):=H_0(X)-H_0(X|Y)$. This is equivalent to the $0$-mutual information~\cite{sason2017arimoto,verdu_alpha_mutual_2015}.

We end this section by presenting the definition of maximin information from non-stochastic information theory~\cite{nair2013nonstochastic}. Consider uvs $X$ and $Y$. Any $x,x'\in\range{X}$ are $\range{X|Y}$-overlap connected if there exists a finite sequence of conditional ranges $\{\range{X|Y(\omega)=y_i}\}_{i=1}^n$ such that $x\in\range{X|Y(\omega)=y_1}$, $x'\in\range{X|Y(\omega)=y_n}$, and $\range{X|Y(\omega)=y_i}\cap \range{X|Y(\omega)=y_{i+1}}\neq \emptyset$ for all $i=1,\dots,n-1$. We say $\mathcal{A}\subseteq\range{X}$ is $\range{X|Y}$-overlap connected if all $x,x'\in\mathcal{A}$ are $\range{X|Y}$-overlap connected. Further, $\mathcal{A},\mathcal{B}\subseteq\range{X}$ are $\range{X|Y}$-overlap isolated if there does not exist $x\in\mathcal{A},x'\in\mathcal{B}$ such that $x,x'$ are $\range{X|Y}$-overlap connected.  An $\range{X|Y}$-overlap partition is a partition of $\range{X}$ such that  each member set is $\range{X|Y}$-overlap connected and any two member sets are $\range{X|Y}$-overlap isolated. There always exists a unique $\range{X|Y}$-overlap partition~\cite{nair2013nonstochastic}, which is denoted by $\range{X|Y}_\star$. The maximin information is $I_\star(X;Y):=\log(|\range{X|Y}_\star|).$
In~\cite{nair2013nonstochastic}, it is proved that  $|\range{X|Y}_\star|=|\range{Y|X}_\star|$ and thus $I_\star(X;Y)=I_\star(Y;X)$. The overlap partition captures common uv~\cite{mahajanrelationship}, an extension of common random variable~\cite{wolf2004zero} to uvs. This relationship explains the relationship between entropy of the common uv, which is equal to the maximin information, and the zero-error capacity~\cite{wolf2004zero,nair2013nonstochastic}.

\section{Information Leakage in Brute-Force Guessing}  \label{sec:info_leakage}

Consider uv $X$ containing sensitive data $U$, which is interpreted as some attribute or feature of $X$ that is computable by some function $g:\range{X}\rightarrow\range{U}$, i.e., $U=g\circ X$. Note that, by construction, $|\range{U}|\leq |\range{X}|$. Let $Y$ be an observable uv that depends on $X$, e.g., $X$ and $Y$ are the input and output, respectively, of a (privacy-preserving) channel.\footnote{The conditional range $\range{Y|X}$ characterizes this channel, which can also be regarded as a non-stochastic privacy-preserving scheme.} These uvs forms a Markov chain $U-X-Y$. An adversary wants to guess $U$ correctly given $Y$. For instance, consider an example in which $X$ captures weight and height of an individual, and $U$ denotes body mass index. In such an example, insurance agencies might be interested in deducing the body mass index of an individual (due to its correlation with heart disease) based on publicly released data $Y$ while they do not have any particular interest in learning an individual's height and weight separately.

We assume that the adversary can guess the value of $U$ in a brute-force trial-and-error manner. That is, the adversary chooses a distinct element $u\in\range{U}$ each time and tests\footnote{We assume that the adversary has access to an oracle that can determine whether $U(\omega)$ is equal to $u$ (for a given $u\in\range{U}$) or not.} whether the actual value $U(\omega)$ equals $u$. The adversary repeats this procedure until the answer is `yes'. We consider the number of trials before the successful guess. Without observations of $Y$, the adversary should try at most $|\range{U}|$ times. However, with access to observation $Y(\omega)=y \in \range{Y}$, the actual value of $U(\omega)$ lies in the conditional range $\range{U|Y(\omega)=y}$ and therefore the maximum number of trials is $|\range{U|Y(\omega)=y}|$. {Since} the number of trials {is} proportional to the inference cost/effort of the adversary, the ratio $|\range{U}|/|\range{U|Y(\omega)=y}|$ captures the reduction in the adversary's maximum cost for guessing $U$ upon the observation $\range{U|Y(\omega)=y}$. This coincide{s} with the definition of the information gain $\log(|\range{U}|/|\range{U|Y(\omega)=y}|)$ in~\cite{kolmogorov1959varepsilon}, where $\log(|\range{U|Y(\omega)=y}|)$ denotes the `combinatorial' conditional entropy. The adversary's reduction in guessing cost can be interpreted as the information gained about uv $U$ from {the} observation $Y(\omega)=y$.

Note that the measure $\log(|\range{U}|/|\range{U|Y(\omega)=y}|)$ is also consistent with the \emph{stochastic brute-force guessing leakage} $H_G(U) - \mathbb{E}_{Y}[H_G(U|Y(\omega)=y)]$
proposed in~\cite[Definition~3]{osia2019privacy} for rvs $U$ and $X$. This measure is based on the guessing entropy\footnote{The guessing entropy $H_G(U)$ denotes the minimum average number of trials for guessing the realization of $U$. This results from the optimal brute-force guessing strategy of the adversary to pick $u_i\in \range{U}$, i.e., the element in $\range{U}$ with the $i$-th largest probability $\mathbb{P}\{U(\omega)=u_i\}$, at the $i$-th trial~\cite{massey1994guessingentropy}. } in~\cite{massey1994guessingentropy} defined as $H_G(U):= \sum_{i=1}^{|\range{U}|} i \mathbb{P}\{U(\omega)=u_i\}$, where $(u_i)_{i=1}^{|\range{U}|}$ are such that $\mathbb{P}\{U(\omega)=u_1\} \geq \mathbb{P}\{U(\omega)=u_2\} \geq \dotsc \geq \mathbb{P}\{U(\omega)=u_{|\range{U}|}\}$.
Similarly, the conditional guessing entropy is $H_{G}(U | Y(\omega)=y) = \sum_{i=1}^{|\range{U|Y(\omega)=y}|} i \mathbb{P}\{U(\omega)=\tilde{u}_i|Y(\omega)=y\}$ for each $y \in \range{Y}$, where $(\tilde{u}_i)_{i=1}^{|\range{U|Y(\omega)=y}|}$ are such that $\mathbb{P}\{U(\omega)=\tilde{u}_1|Y(\omega)=y\} \geq \mathbb{P}\{U(\omega)=\tilde{u}_2|Y(\omega)=y\} \geq \dotsc \geq \mathbb{P}\{U(\omega)=\tilde{u}_{|\range{U|Y(\omega)=y}|}|Y(\omega)=y\}$. When there is no $\sigma$-field or probability measure over $\range{U}$, $H_{G}(U)$ and $H_G(U|Y(\omega)=y)$ reduce to the prior guessing cost $\log(|\range{U}|)$ and posterior guessing cost $\log(|\range{U|Y(\omega)=y}|)$, respectively, {by} replacing the {expectation} with the worst-case. To quantify the \emph{non-stochastic brute-force guessing leakage}, we consider the difference between $\log(|\range{U}|)$ and the minimum guessing cost $\min_{y \in \range{Y}}\log(|\range{U|Y(\omega)=y}|)$ as follows.

\begin{definition}[Non-Stochastic Brute-force Guessing Leakage]
For a given uv $U$, the non-stochastic leakage from $U$ to $Y$ is
\begin{align*}
\Leak(U\rightarrow Y)=&\log
\left(\frac{|\range{U}|}{\displaystyle\min_{y\in \range{Y}} |\range{U|Y(\omega)=y}|} \right)
\\
=&
 \max_{y\in \range{Y}} \log
\left(\frac{|\range{U}|}{|\range{U|Y(\omega)=y}|} \right).
\end{align*}
\end{definition}

The measure $\Leak(U \rightarrow Y)$ quantifies the maximum reduction in the guessing cost of the adversary after observing $Y$, which indicates the most information gained by the adversary in the sense of~\cite{kolmogorov1959varepsilon}. This measure has been previously used as the non-stochastic information leakage in~\cite{8662687,ding2019developing} for privacy analysis, e.g., in the case of $k$-anonymity~\cite{8662687}. Hence, this definition provides an operative meaning to the non-stochastic information leakage and can be used as its interpretation for privacy analysis.

In the following proposition, we show that non-stochastic leakage satisfies the data-processing inequality. This implies that, for a given uv $X$ and a specified attribute $U$ of $X$, the leakage is non-increasing along cascading channels $\range{Y|X}$ and $\range{Z|Y}$. This is in line with axiom (R3.a) of an operational notion of information leakage in \cite{issa2018operational}. This is an important requirement as it shows that a curator does not need to worry about an increased risk {incurred by any post processing} after releasing outputs.

\begin{proposition}[Data Processing Inequality] If Markov chain $U-X-Y-Z$ holds, $\Leak(U\rightarrow Z)\leq \Leak(U\rightarrow Y)$.
	\label{prop:DPI_1}
\end{proposition}

\begin{IEEEproof} Note that
	\begin{align*}
	\range{U|Z(\omega)=z}
	&=\bigcup_{y\in\range{Y}} \range{U|Z(\omega)=z,Y(\omega)=y}\\
	&=\bigcup_{y\in\range{Y}:(y,z)\in\range{Y,Z}} \range{U|Y(\omega)=y},
	\end{align*}
	where the last equality follows from that $U-Y-Z$ is  a Markov chain, i.e., $U$ and $Z$ are unrelated given $Y$.
	Notice that $y\in\range{Y}$ and $(y,z)\in\range{Y,Z}$ implies that $y\in\range{Y|Z(\omega)=z}$. As a result,
	\begin{align} \label{eqn:proof:1}
	\range{U|Z(\omega)=z}
	&=\bigcup_{y\in\range{Y|Z(\omega)=z}} \range{U|Y(\omega)=y}.
	\end{align}
	Let $z^* \in \argmin_{z\in\range{Z}} |\range{U|Z(\omega)=z}|$. For any $y^*\in\range{Y|Z(\omega)=z^*}$, $\range{U|Y(\omega)=y^*} \subseteq \range{U|Z(\omega)=z^*} $ because of~\eqref{eqn:proof:1}. Hence,
	\begin{align*}
	\min_{y\in\range{Y}} |\range{U|Y(\omega)=y}|
	&\leq |\range{U|Y(\omega)=y^*}|\\
	&\leq |\range{U|Z(\omega)=z^*}|\\
	&=\min_{z\in\range{Z}}|\range{U|Z(\omega)=z}|,
	\end{align*}
	which, because of the monotonicity of the logarithm, gives rise to the inequality $\Leak(U\rightarrow Z)\leq \Leak(U\rightarrow Y)$.
\end{IEEEproof}

The following result shows that the non-stochastic brute-force guessing leakage is a measure of relatedness between two uvs. In fact, the leakage is equal to zero if two uvs are unrelated. Evidently, the most private case arises from ensuring that $X$ and $Y$ are unrelated. In this case, the realizations of $Y$ do not provide any useful information about $X$ or its derivatives, e.g., $U$.  This is again in line with axiom (R3.b) of an operational notion of information leakage~\cite{issa2018operational}.

\begin{proposition}[Bounding Leakage]
	\label{prop:Bounding_Leakage}
	 $ \Leak(U\rightarrow Y)\geq 0$ with equality if $X$ and $Y$ are unrelated.
\end{proposition}
\begin{IEEEproof} The inequality follows from that $\range{U|Y(\omega)=y}\subseteq \range{U}$ and, as a a result, $|\range{U}|/|\range{U|Y(\omega)=y}|\geq 1$. If $X$ and $Y$ are unrelated, $U$ and $Y$ are unrelated too. Therefore, $\range{U|Y(\omega)=y}=\range{U}$. This shows that $\Leak(U\rightarrow Y)=0$.
\end{IEEEproof}

For the Markov chain $U-X-Y$, the measure $\Leak(U \rightarrow Y)$ can be used to quantify the non-stochastic brute-force guessing leakage if we know attribute $U$ of $X$ that is targeted by the adversary. However, there are some real-world situations that we do not know \emph{a priori} the intention of the adversary, i.e., the attribute $U$ of $X$ that the adversary is trying to infer. In some cases, more than one user may observe $Y$ and each user might be interested in guessing/estimating a different attribute of $X$. In these situations, it is required to consider the brute-force guessing leakage $\Leak(U \rightarrow Y)$ when the attribute $U$ varies. Therefore, we need to define a {maximal non-stochastic} guessing leakage. This is in-line with axiom (R2) in~\cite{issa2018operational}. We consider such situations in the next section.

\begin{definition}[Maximal Non-Stochastic Brute-Force Guessing Leakage]
The maximal non-stochastic leakage from $X$ to $Y$ is defined as
\begin{align}\label{eqn:maximal_leakage}
\LeakS(X\rightarrow Y)=&\sup_{U \colon U-X-Y}\Leak(U\rightarrow Y),
\end{align}
where the supremum is taken over all {functions} $g:\range{X}\rightarrow\range{U}$ with $\range{U}$ containing finite arbitrary alphabets.
\end{definition}

Now, we can show that maximal non-stochastic leakage admits axiom (R3) in the axiomatic approach to operational information leakage in~\cite{issa2018operational}. That is, maximal non-stochastic leakage satisfies data processing inequality (post processing does not increase leakage), independence property (statistically independent outputs result in zero leakage), and additivity.

\begin{proposition}[Properties of Maximal Leakage] \label{prop:axiom}
The following holds:
\begin{itemize}
\item[a)] $\LeakS(X\rightarrow Y)\geq 0$;
\item[b)] $\LeakS(X\rightarrow Y)=0$ if and only if $X$ is unrelated to $Y$;
\item[c)] $\LeakS(X\rightarrow Y)\leq H_0(X)$ with the equality if $Y = X$;
\item[d)] 
$\LeakS(X\rightarrow Z)\leq \LeakS(X\rightarrow Y)$ if Markov chain $X-Y-Z$ holds;
\item[e)] If $(X_i,Y_i)$, $\forall i$, are unrelated, i.e., $(X_i,Y_i)$ and $(X_{i'},Y_{i'})$ are unrelated $\forall i\neq i'$, then $\LeakS((X_1,\dots,X_n)\rightarrow (Y_1,\dots,Y_n))
    =\sum_{i=1}^n\Leak(X_i\rightarrow Y_i)$.   \hfill \IEEEQED
\end{itemize}
\end{proposition}

Now, we are ready to present a formula for computing the maximal non-stochastic leakage. This is done in the next proposition.

\begin{proposition}[Computing Maximal Leakage] \label{prop:Bounds_on_Maximal_Leakage}
$\LeakS(X\rightarrow Y)=\log(|\range{X}| - \min_{y\in\range{Y}} |\range{X|Y(\omega)=y}| + 1)$.
\end{proposition}

The proofs of Propositions~\ref{prop:axiom} and \ref{prop:Bounds_on_Maximal_Leakage} are in Appendix~\ref{app:prop:axiom} and \ref{app:prop:Bounds_on_Maximal_Leakage}, respectively.
In Appendix~\ref{app:prop:Bounds_on_Maximal_Leakage}, The function $g$ in~\eqref{eq:WorstCaseAttri} constructs the most vulnerable attribute $U$ of uv $X$, which is determined by any ${y^* \in }\argmin_{y\in\range{Y}}|\range{X|Y(\omega)=y}|$.
The following example shows that the maximal non-stochastic brute-force guessing leakage is incurred when the function $g$ is highly corrected with the mapping $\range{Y|X}$.

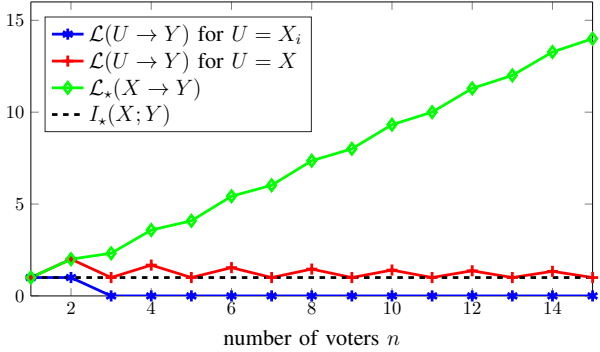
\begin{figure}[t]
	\centering
	\scalebox{0.7}{
%

%
\begin{tikzpicture}

\begin{axis}[
width=4.2in,
height=2.2in,
scale only axis,
separate axis lines,
xmin=1,
xmax=15,
xlabel = {\large number of voters $n$},
ymin=0,
ymax=16,
legend style={at={(0.5,0.95)},draw=darkgray!60!black,fill=white,legend cell align=left}
]

\addplot [
color=blue,
solid,
line width=1.5pt,
mark=asterisk,
mark options={solid},
mark size=3.0pt,
]
table[row sep=crcr]{
1 1\\
2 1\\
3 0\\
4 0\\
5 0\\
6 0\\
7 0\\
8 0\\
9 0\\
10 0\\
11 0\\
12 0\\
13 0\\
14 0\\
15 0\\
};
\addlegendentry{\large $\Leak(U\rightarrow Y)$ for $U = X_i$};

\addplot [
color = red,
solid,
line width=1.5pt,
mark=+,
mark size=3.0pt,
]
table[row sep=crcr]{
1 1\\
2 2\\
3 1\\
4 1.67807190511264\\
5 1\\
6 1.5405683813627\\
7 1\\
8 1.46084118889197\\
9 1\\
10 1.40754296273192\\
11 1\\
12 1.36882294429602\\
13 1\\
14 1.33911272969743\\
15 1\\
};
\addlegendentry{\large $\Leak(U\rightarrow Y)$ for $U = X$};

\addplot [
color=green,
solid,
line width=1.5pt,
mark=diamond,
mark size=3.0pt,
]
table[row sep=crcr]{
	1 1\\
	2 2\\
	3 2.32192809488736\\
	4 3.58496250072116\\
	5 4.08746284125034\\
	6 5.4262647547021\\
	7 6.02236781302845\\
	8 7.35755200461808\\
	9 8.00562454919388\\
	10 9.31967212094699\\
	11 10.0014081943928\\
	12 11.2940463132715\\
	13 12.0003521774803\\
	14 13.2745237550067\\
	15 14.0000880524301\\
};
\addlegendentry{\large $\LeakS(X \rightarrow Y)$};

\addplot [
color=black,
dashed,
line width=1.5pt,
]
table[row sep=crcr]{
1 1\\
2 1\\
3 1\\
4 1\\
5 1\\
6 1\\
7 1\\
8 1\\
9 1\\
10 1\\
11 1\\
12 1\\
13 1\\
14 1\\
15 1\\
};
\addlegendentry{\large $I_\star(X;Y)$};

\end{axis}
\end{tikzpicture}%
}
	\caption{The non-stochastic brute-force guessing leakage $\Leak(U \rightarrow Y)$ and the maximal non-stochastic brute-force guessing leakage $\LeakS(X \rightarrow Y)$ vs. the number of voters $n$ in the majority vote in Example~\ref{ex:VoteMain}. Two cases for $U$ in $\Leak(U \rightarrow Y)$ are considered. First, $U = X_i$ when the adversary tries to exhaustively guessing user $i$'s vote. Then, $U = X = (X_i \colon i \in \Set{1,\dotsc, n})$ when the adversary tries to exhaustively guess all votes. }
	\label{fig:LeakVote}
\end{figure}

 \begin{example}[Majority Vote]  \label{ex:VoteMain}
 Let uncertain variable $X_i:\Omega\rightarrow\range{X_i}=\{0\text{-``no/disagree''},1\text{-``yes/agree''}\}$ denotes the vote of individual $i$. Assume that there are (finite) $n \in \mathbb{N}$ voters. Denote $X(\omega) = (X_i(\omega) \colon i \in \Set{1,\dotsc,n})$ the voting result. Let $Y(\omega)=\mathfrak{q}(X(\omega))$ be the majority vote function such that
    \begin{equation}  \label{eq:MajorVote}
    \mathfrak{q}(x) = \begin{cases} 0, & \displaystyle \sum_{i= 1}^n x_i < \frac{n}{2}, \\ 1, & \displaystyle \sum_{i= 1}^n x_i \geq \frac{n}{2}, \end{cases}
    \end{equation}
for $x = (x_i \colon i \in \Set{1,\dotsc,n}) \in \Set{0,1}^n$.
 Consider the following two attributes $U$.
 In this example, we have set the base of logarithm to $2$.

Let $U=X_i$, i.e., the adversary is interested in an individual's vote. It can be seen that
 \begin{align*}
     \range{U|Y(\omega)=0}
     &=\begin{cases}
     \{0\}, & n=1, \\
     \{0\}, & n=2, \\
     \{0,1\}, & n\geq 3,
     \end{cases}\\
     \range{U|Y(\omega)=1}
     &=\begin{cases}
     \{0\}, & n=1, \\
     \{0,1\}, & n=2, \\
     \{0,1\}, & n\geq 3.
     \end{cases}
 \end{align*}
Therefore,
 \begin{align}
     \Leak(U\rightarrow Y)=& \nonumber
     \log_2\hspace{-.02in}
 \left(\frac{|\range{U}|}{\displaystyle\min_{y\in \range{Y}} |\range{U|Y(\omega)=y}|} \right)\\
 &=
 \begin{cases}
     1, & n=1, \\
     1, & n=2, \\
     0, & n\geq 3.
     \end{cases}\label{eq:USingle}
 \end{align}
 The information leakage about the vote of each individual is zero if there are more than three voters. This is practically why democracy with secrete/confidential ballot is privacy preserving~\cite{bernhard2017public}.

Let $U = X$, i.e., the adversary is interested in figuring out all the votes. We have
  \begin{align*}
      |\range{X|Y(\omega)=0}|
      &=\left|\left\{ x \in \Set{0,1}^n \colon \sum_{i = 1}^{n} x_i < \frac{n}{2} \right\}\right| \\
      &= \sum_{k=0}^{n_0} {n \choose k}, \\
      |\range{X|Y(\omega)=1}|
      &= \left|\left\{ x \in \Set{0,1}^n \colon \sum_{i = 1}^{n} x_i \geq \frac{n}{2} \right\}\right| \\
      &=\sum_{k=n_0+1}^{n} {n \choose k}.
  \end{align*}
where
 \begin{align*}
     n_0
     = \left\lceil  \frac{n}{2} \right\rceil - 1 =
     \begin{cases}
         \displaystyle \frac{n-1}{2}, & n\mbox{ is odd},\\
         \displaystyle \frac{n}{2}-1, & n\mbox{ is even}.
     \end{cases}
 \end{align*}
Then, $|\range{X|Y(\omega)=0}| - |\range{X|Y(\omega)=1}|<0$ if $n$ is odd and $|\range{X|Y(\omega)=0}| - |\range{X|Y(\omega)=1}|=0$ if $n$ is even.
Therefore,
 \begin{align}
     \Leak(U\rightarrow Y) &= \Leak(X\rightarrow Y)  \nonumber \\
     &= \log_2 \left(\frac{|\range{X}|}{\min_{y\in \{0,1\}} |\range{X|Y(\omega)=y}|} \right) \nonumber \\
     &= \log_2 \left(\frac{2^n}{ |\range{X|Y(\omega)=0}|} \right) \nonumber \\
     &= n-\log_2 \left(\sum_{k=0}^{n_0} {n \choose k}\right). \label{eq:UAll}
 \end{align}
For odd $n$,~\cite[p.\,167]{graham1994concrete} implies that
\begin{align*}
    \sum_{k=0}^{n_0} {n \choose k}=\sum_{k=0}^{n_0} {2n_0+1 \choose k}=2^{2n_0},
\end{align*}
and as a result $\Leak(U\rightarrow Y)=n-2n_0=1.$
For even $n$,
\begin{align*}
    \sum_{k=0}^{n_0} {n \choose k}
    &=\sum_{k=0}^{n_0} {2n_0+2 \choose k}\\
    & \geq \sum_{k=0}^{n_0} {2n_0+1 \choose k}\\
    & =2^{2n_0},
\end{align*}
and thus $\Leak(U\rightarrow Y)\leq n-2n_0=2.$ These derivations show that $\Leak(U\rightarrow Y) \in [0,2]$.
That is, $\Leak(U\rightarrow Y)$ is independent of $n$, even if $|\range{U}| = 2^n$ exponentially grows with $n$.

However, if we do not know the privacy-intrusive adversary' intention, i.e., which attribute $U$ is targeted, we should prepare for the worst-case privacy leakage.
To see the most vulnerable attribute, we follow \eqref{eq:WorstCaseAttri} to construct function
\begin{equation} \label{eq:HPF}
    g(x)=
        \begin{cases}
            u^*, & x \in \range{X|Y(\omega)=0},\\
            x, & x \in \range{X|Y(\omega)=1}.
        \end{cases}
\end{equation}
The maximal (worst-case) non-stochastic brute-force leakage
\begin{align*}
    \LeakS(X\rightarrow Y) &= \log_2 (|\range{X}| -\range{X|Y(\omega) = 0} + 1)\\
                           &= \log_2 \Big( 2^n - \sum_{k=0}^{n_o} {n \choose k} + 1 \Big)\\
                           &= \log_2 (|\range{X | Y(\omega)=1}| + 1),
\end{align*}
which is increasing in $n$.
Note that $g$ in \eqref{eq:HPF} is equivalent with the exact voting results only when no less than half of the users vote ``yes/agree''.
This function is highly related to $\mathfrak{q}$, the majority vote function that determines $\range{Y|X}$, and therefore generates the most vulnerable attribute of $X$.
Here, $g$ is only related to but is not exactly $\range{Y|X}$. We will show in Section~\ref{sec:relationship} that the attribute determined exactly by $\mathfrak{q}$ maximizes another worst-case non-stochastic guessing leakage, the maximin information.

Fig.~\ref{fig:LeakVote} shows the non-stochastic brute-force guessing leakage $\Leak(U \rightarrow Y)$ for $U$ being an individual's vote and all votes, and the maximal non-stochastic brute-force leakage $\LeakS(X \rightarrow Y)$  versus the number of voters. Evidently, the most destructive malicious inference is not exhaustively guessing all individuals' votes, but inferring an uncertainty that is highly related to the majority voting results.
\end{example}

\begin{remark}
$\LeakS(X\rightarrow Y)$ is not symmetric in general. In most cases, $\LeakS(X\rightarrow Y) \neq \LeakS(Y\rightarrow X)$. For example, for uvs $X$ and $Y$ with joint range $\range{X,Y}=\{(x_1,y_1),(x_2,y_1),(x_3,y_2)\}$, we have  $\LeakS(X\rightarrow Y)=\log(3)\neq \log(2)=\LeakS(Y\rightarrow X)$.
Most of the existing information leakage measures are asymmetric, e.g., the (stochastic) maximal leakage (Sibson mutual information) \cite{issa2018operational}, $\alpha$-leakage \cite{liao2018tunable},  \cite{RenyiDP}.
That is, the quantity of information depends on the direction of the data flow.
\end{remark}


\section{Non-Stochastic Identifiability} \label{sec:privacy}
We define non-stochastic identifiability by requiring that the ratio of the cardinality of the set of compatible realization of uv $X$ with access to the measurements of uv $Y$ over the cardinality of the set of compatible realization of uv $X$ without this auxiliary information is lower bounded by an exponential of the privacy budget. This implies that access to the realizations of $Y$ does not significantly reduce the cardinality of the set of possibilities that must be tested for guessing the realization of $X$.
This definition is consistent with stochastic identifiability in~\cite[Definition~2]{wang2016relation}, \cite[Definition~4]{lee2012differential} and \cite[Definition~3.1]{Membership2013}, which require that the posterior distribution (instead of the conditional range) to remain similar with and without access to the measurements.
For attribute $U$, the prior range $\range{U}$ and posterior range $\range{U|Y}$ denote the adversary's uncertainty about $U$ before and after observing $Y$, respectively, and their logarithmic difference captures non-probabilistic identifiability.

\begin{definition}[Non-Stochastic Identifiability] For $\epsilon > 0$, a mapping $\mathfrak{M}$ that generates $Y=\mathfrak{M}\circ X$ is $\epsilon$-identifiable on attribute $U$ if
\begin{align} \label{eq:NSInd}
	 \Leak(U \rightarrow Y) = \log \left(\frac{\range{U}}{\min_{y \in \range{Y}}\range{U | Y(\omega) = y}}\right) \leq \epsilon,
\end{align}
for all $y \in \range{Y}$.
\end{definition}
In this definition, \eqref{eq:NSInd} ensures $\log ({\range{U}}/{\range{U | Y(\omega) = y}}) \leq \epsilon, \forall y \in \range{Y}$, which corresponds to the membership privacy $\log({p(u|y)}/{p(u)}) \leq \epsilon, \forall u, y$ in \cite[Definition~3.1]{Membership2013} for the probability space.
As in stochastic notions of privacy, we refer to $\epsilon$ in the non-stochastic identifiability as {the} \emph{privacy budget}, i.e., by decreasing $\epsilon$, we ensure a higher level of privacy (cf., differential privacy~\cite{dwork2006calibrating} and identifiability~\cite{wang2016relation}). This is intuitively because, by decreasing the privacy budget, the size of the set $\range{U|Y(\omega)=y}$ increases and thus guessing the actual realization of uv $U$ becomes harder.

\subsection{Identifiability and $\Leak(X \rightarrow Y)$}

Identifiability is closely related the maximal non-stochastic brute-force guessing leakage.
It can be seen from the corresponding definitions that  $\Leak(X \rightarrow Y)$ and $\Leak_* (X \rightarrow Y)$ are both decreasing functions of $\min_{y \in \range{Y}}  \range{X | Y(\omega) = y}$. Also, for the natural logarithm, they can express in terms of each other by
\begin{equation} \label{eq:LeakLeakS}
	e^{\LeakS(X \rightarrow Y)}  + |\range{X}| e^{-\Leak(X \rightarrow Y)} = |\range{X}| + 1.
\end{equation}
Therefore, $\LeakS(X \rightarrow Y)$ is increasing in $\Leak(X \rightarrow Y)$ and \textit{vice versa}; see
more on this relationship in Appendix~\ref{app:eq:LeakLeakS}. Thus, imposing a budge on either measure necessarily bounds the other.

\begin{corollary} \label{cor:ident}
For any $\epsilon$-identifiable mapping $\mathfrak{M}$, $\LeakS(X\rightarrow Y)\leq \log(|\range{X}|(1-e^{-\epsilon}) + 1)$.
\end{corollary}

\begin{IEEEproof}
The proof follows from \eqref{eq:LeakLeakS} that
$\LeakS(X\rightarrow Y)
    =\log(|\range{X}| - \min_{y\in\range{Y}} |\range{X|Y(\omega)=y}| + 1)
    \leq \log(|\range{X}|(1-e^{-\epsilon}) + 1)$.
\end{IEEEproof}

Corollary~\ref{cor:ident} shows that, as expected, the maximal non-stochastic brute-force guessing leakage $\LeakS(X\rightarrow Y)$ goes to zero as the privacy budget approaches zero. By increasing the privacy budget $\epsilon$, however, we increase the bound on the maximal non-stochastic brute-force guessing leakage $\LeakS(X\rightarrow Y)$ and therefore more private information could be potentially leaked.

\section{Brute-Force to One-Shot Guess}
\label{sec:relationship}

In the previous sections, we considered a brute-force guessing adversary that can potentially check all the possibilities of the private information in $\range{U|Y(\omega)=y}$ that are compatible with the available outputs $Y(\omega)=y$ of the channel $\range{Y|X}$ to find the actual private realization. In this section, we restrict ourselves to one-shot guesses. We first analyze the non-stochastic case and its relationship with the non-stochastic brute-force guessing.

\subsection{Non-Stochastic One-Shot Guessing}
Let us consider an adversary with only a single opportunity for guessing the private realization of uv $U$ by observing the realization of uv $Y$. For instance, consider the problem of guessing a person's password based on side-channel information (e.g., inter-keystroke delay as in~\cite{issa2018operational}) while the system locks immediately after one wrong guess. Therefore, the adversary is interested in finding the largest amount of information that can be deduced correctly with one guess. This happens when $|\range{U|Y(\omega)=1}|=1$ for all $y\in\range{Y}$. In the following proposition, we show that the maximum information is the largest amount of information can be leaked to such an adversary. We further relate this notion of leakage to maximal non-stochastic leakage with brute-force guessing.

\begin{proposition}[Maximal Leakage Bounds Maximin Information] \label{prop:maximin}
For uvs $X$ and $Y$,
\begin{align*}
    I_\star(X;Y)=
    \hspace{-.1in}
    \sup_{
    \scriptsize
    \begin{array}{c}
    \scriptsize U\colon U-X-Y, \\
    \scriptsize |\range{U|Y(\omega)=y}|=1, \\
    \forall y\in\range{Y}
    \end{array}
    }
    \hspace{-.1in}\Leak(U\rightarrow Y)
    \leq
    \LeakS(X\rightarrow Y),
\end{align*}
where the supremum is taken over  all $g:\range{X}\rightarrow\range{U}$ such that $|\{g(x) \colon x \in \range{X | Y(\omega)=y}\}| =  |\range{U|Y(\omega)=y}|=1$.
\end{proposition}

\begin{IEEEproof}
	The second inequality trivially follows from that increasing the search domain of the supremum operator results in a larger value. Therefore, we only focus on the first inequality. Note that $|\range{U|Y(\omega)=y}|=1$ implies that there exists $f$ such that $U=f(Y)$. Therefore, $U=f(Y)=g(X)$. Following Lemma 1 in~\cite{wolf2004zero}, we know that there exists a function $h$ such that $U=h(X\wedge Y)$, where $X\wedge Y$ is the common variable in the sense of~\cite{wolf2004zero} defined for uncertain variables (instead of random variables) following the approach of~\cite{mahajanrelationship}. Therefore,
	$\Leak(U\rightarrow Y)=H_0(U)\leq H_0(X\wedge Y)=I_\star(X;Y).$
	Since this inequality holds for all $U$ such that $\range{U|Y(\omega)=y}=1$, we get
	\begin{align*}
	    \sup_{
    \scriptsize
    \begin{array}{c}
    \scriptsize U\colon U-X-Y, \\
    \scriptsize |\range{U|Y(\omega)=y}|=1, \\
    \forall y\in\range{Y}
    \end{array}
    }\Leak(U\rightarrow Y)\leq
	    I_\star(X;Y).
	\end{align*}
 	{On the other hand, for $U^* = X \wedge Y$,
	\begin{align*}
	    \sup_{
    \scriptsize
    \begin{array}{c}
    \scriptsize U\colon U-X-Y, \\
    \scriptsize |\range{U|Y(\omega)=y}|=1, \\
    \forall y\in\range{Y}
    \end{array}
    }\hspace{-.2in}\Leak(U\rightarrow Y)
    \geq \Leak(U^*\rightarrow Y)=I_\star(X;Y).
	\end{align*}
	}
	Combining these inequalities concludes the proof.
\end{IEEEproof}


It was recently proved that the maximin information $I_\star(X;Y)$ is equal to the entropy of the common uncertain variable between uvs $X$ and $Y$~\cite{9457645}. Common uncertain variable is defined similar to common random variable in~\cite{wolf2004zero} and is the largest uncertain variable that can be directly computed using both $X$ and $Y$. The ``largest'' uncertain variable refers to that any other uncertain variable that can be directly computed using both $X$ and $Y$ can also be written in terms of the common uncertain variable. The definition of common uncertain variable paved the way to define perfect privacy~\cite{9457645}. Given this relationship, Proposition~\ref{prop:maximin} demonstrates the relationship between common uncertain variable and non-stochastic one-shot guessing.

 \begin{example}[Majority Vote (Cont.)]  \label{ex:VoteAux}
 	For the majority voting function in Example~\ref{ex:VoteMain}, the maximin information is
 	\begin{align}
 		&\range{U_\star|X(\omega) = x}  \nonumber \\
 		& \ = \argmax_{U} \Big\{ \Leak(U \rightarrow Y) \colon U-X-Y, \range{U|Y(\omega)=y}=1 \Big\} \nonumber\\
 		& \ = \begin{cases}
 				\Set{a} & x \in \range{X | Y(\omega) = 0}\\
 				\Set{b} & x \in \range{X |Y(\omega)= 1}
 			\end{cases} \label{eq:IStarEx}
 	\end{align}
 This is exactly the majority vote function $\mathfrak{q}$ in \eqref{eq:MajorVote}. Therefore, an adversary can perfectly estimate $U^*$ by observing $Y$. This is however not a privacy breach as $Y$ is already disclosed. Also, \eqref{eq:IStarEx} is independent of $n$ and thus the maximin information is $I_\star(X;Y) = \log_2 (2)  = 1$ for all $n \in \mathbb{N}$. See Fig.~\ref{fig:LeakVote}.
 \end{example}

\begin{remark}[Relationship with Zero-Error Capacity]
Following Proposition~\ref{prop:maximin} and~\cite{nair2013nonstochastic}, the zero-error capacity of any memoryless uncertain channel satisfies $C_0=\sup_{\range{X}\subseteq\mathbb{X}}I_\star (X;Y)\leq  \sup_{\range{X}\subseteq\mathbb{X}} \LeakS(X\rightarrow Y).$
Therefore, based on Corollary~\ref{cor:ident}, the zero-error capacity of any memoryless $\epsilon$-identifiable channel is upper bounded by $\log(|\mathbb{X}|(1-2^{-\epsilon}) + 1)$, where $|\mathbb{X}|$ is the number of the input alphabets.
This constraints dynamical systems that can be estimated or stabilized through privacy-preserving communication channels~\cite{nair2013nonstochastic, matveev2007shannon}.
\end{remark}

In the next subsection, we consider one-shot guessing in the stochastic sense of~\cite{issa2018operational} and investigate its relationship with the maximal non-stochastic leakage with brute-force guessing.

\subsection{Maximal Stochastic Leakage}
We can recreate the stochastic framework for information leakage in~\cite{issa2018operational} by endowing all the uncertain variables in this paper with a measure.

\begin{definition}[Stochastic Maximal Leakage]
	For jointly distributed rvs $X$ and $Y$, the maximal stochastic leakage from $X$ to $Y$ is given by
	\begin{align*}
	\widetilde{\LeakS}(X&\rightarrow Y)\\
	&=
	\sup_{U\colon U-X-Y} \log\left( \frac{\displaystyle\mathbb{E}\left\{\max_{u \in \range{U}} \mathbb{P}\{U=u|Y=y\}\right\}}{\displaystyle\max_{u\in\range{U}}\mathbb{P}\{U=u\}}\right),
	\end{align*}
	where supremum is taken over all {random variables (rvs)} $U$ taking values in finite arbitrary alphabets. It was shown in~\cite{issa2018operational} that
	\begin{align*}
	\widetilde{\LeakS}(X\rightarrow Y)
	&=\log\left(\sum_{y\in\range{Y}}\max_{x\in\range{X}}\mathbb{P}\{Y=y|X=x\}\right)\\
	&=I_{\infty} (X;Y),
	\end{align*}
	where $I_{\infty}$ is the Sibson mutual information $I_{\alpha}$ in the order $\alpha \rightarrow \infty$~\cite{sibson_information_1969, verdu_alpha_mutual_2015}. Note the fact that $\{x\colon \mathbb{P}\{X=x\}>0\} = \range{X}$.
\end{definition}

\begin{figure*}[t]
	\subfigure{\scalebox{0.7}{
%
%
\begin{tikzpicture}

\begin{axis}[%
width=4.2in,
height=2.2in,
scale only axis,
xmin=1,
xmax=29,
xlabel = {\large utility loss/maximum distortion: $\max_{y \in \range{Y}}  |y-\hat{c}(y)| $},
ymin=8.06,
ymax=8.21,
ylabel = {\large privacy leakage: $\LeakS(X \rightarrow \hat{Y})$},
legend style={at={(0.98,0.95)},draw=darkgray!60!black,fill=white,legend cell align=left}
]

\addplot [
color=blue,
solid,
line width=1.5pt,
mark=asterisk,
mark options={solid},
mark size=3.0pt,
]
table[row sep=crcr]{
1 8.19475685442225\\
2 8.19475685442225\\
4 8.19475685442225\\
5 8.19475685442225\\
7 8.19475685442225\\
8 8.18982455888002\\
10 8.18982455888002\\
11 8.15987133677839\\
13 8.15987133677839\\
14 8.15987133677839\\
16 8.15987133677839\\
17 8.15987133677839\\
19 8.13442632022093\\
20 8.13442632022093\\
22 8.13442632022093\\
23 8.13442632022093\\
25 8.13442632022093\\
26 8.13442632022093\\
28 8.06069593168755\\
29 8.06069593168755\\
};
\addlegendentry{\large $Y=$``blood pressure"};

\addplot [
color=red,
solid,
line width=1.5pt,
mark=diamond,
mark options={solid},
mark size=3.0pt,
]
table[row sep=crcr]{
1 8.19475685442225\\
2 8.19475685442225\\
4 8.19475685442225\\
5 8.18982455888002\\
7 8.18487534290828\\
8 8.17990909001493\\
10 8.17990909001493\\
11 8.13442632022093\\
13 8.13442632022093\\
14 8.13442632022093\\
16 8.09275714091985\\
17 8.09275714091985\\
19 8.09275714091985\\
20 8.09275714091985\\
22 8.07146236255662\\
23 8.07146236255662\\
25 8.07146236255662\\
26 8.07146236255662\\
28 8.07146236255662\\
29 8.07146236255662\\
};
\addlegendentry{\large $Y=$``cholesterol"};
\end{axis}
\end{tikzpicture}%
}} \qquad
	\subfigure{\scalebox{0.7}{
%
%
\begin{tikzpicture}

\begin{axis}[%
width=4.2in,
height=2.2in,
scale only axis,
xmin=1,
xmax=29,
xlabel = {\large utility loss/maximum distortion: $\max_{y \in \range{Y}}  |y-\hat{c}(y)| $},
ymin=2,
ymax=8,
ylabel = {\large privacy leakage: $I_\star (X ; Y)$},
legend style={at={(0.98,0.95)},draw=darkgray!60!black,fill=white,legend cell align=left}
]

\addplot [
color=blue,
solid,
line width=1.5pt,
mark=asterisk,
mark options={solid},
mark size=3.0pt,
]
table[row sep=crcr]{
1 5\\
2 4.39231742277876\\
4 3.90689059560852\\
5 3.58496250072116\\
7 3.32192809488736\\
8 3.16992500144231\\
10 3\\
11 2.8073549220576\\
13 2.58496250072116\\
14 2.58496250072116\\
16 2.58496250072116\\
17 2.32192809488736\\
19 2.32192809488736\\
20 2\\
22 2\\
23 2\\
25 2\\
26 2\\
28 2\\
29 2\\
};
\addlegendentry{\large $Y=$"blood pressure"};

\addplot [
color=red,
solid,
line width=1.5pt,
mark=diamond,
mark options={solid},
mark size=3.0pt,
]
table[row sep=crcr]{
1 7.25738784269265\\
2 6.04439411935845\\
4 5.49185309632967\\
5 5.12928301694497\\
7 4.90689059560852\\
8 4.58496250072116\\
10 4.52356195605701\\
11 4.32192809488736\\
13 4.16992500144231\\
14 4\\
16 3.90689059560852\\
17 3.8073549220576\\
19 3.70043971814109\\
20 3.70043971814109\\
22 3.58496250072116\\
23 3.4594316186373\\
25 3.4594316186373\\
26 3.32192809488736\\
28 3.32192809488736\\
29 3.16992500144231\\
};
\addlegendentry{\large $Y=$"cholesterol"};

\end{axis}
\end{tikzpicture}%
}}
	\caption{The privacy-utility tradeoff: The maximal non-stochastic brute-force guessing leakage $\LeakS(X \rightarrow \hat{Y})$ (left) and maximin information $I_\star(X;\hat{Y})$ (right) versus the maximal distortion between $Y$ and $\hat{Y}$. The plots are obtained for the experiment in Section~\ref{sec:experiment} on the heart disease dataset from the UCI machine learning repository \cite{UCI2007}. Here, $\hat{Y}$ denotes the quantized $Y$ by~\eqref{eq:Quantizer}, where the step size $\delta$ varies between $1$ a and $50$. The maximum distortion refers to the largest $\ell_1$ distance between the actual value $y$ and the centroid $\hat{c}(y)$. }
	\label{fig:PUTLeakS}
\end{figure*}

It is shown in \cite[Lemma~1]{issa2018operational} that the worst-case maximal stochastic leakage occurs when the probability becomes deterministic:
\begin{align*}
	\sup_{p(y|x)} \widetilde{\LeakS}(X\rightarrow Y)
	& =  \min \Set{ H_0(X), H_0(Y)} \\
	& = \min \Set{ |\range{X}|, |\range{Y}| }
\end{align*}

In non-stochastic case, we have
$ \sup_{\range{Y|X}} \LeakS (X \rightarrow Y) = H_0(X) $
and
$ \sup_{\range{Y|X}}I_\star(X;Y) = \min \{ H_0(X), H_0(Y) \}$.
Therefore,
\begin{align}
	\sup_{p(y|x)} \widetilde{\LeakS}(X\rightarrow Y)
	&=  \sup_{\range{Y|X}}I_\star(X;Y) \nonumber  \\
	&\leq \sup_{\range{Y|X}} \LeakS (X \rightarrow Y) . \label{eq:RelationLeak}
\end{align}
The equality holds when $\range{X} \leq \range{Y}$.

\begin{figure}[t]
	\centering
	\scalebox{0.7}{
%
%
\begin{tikzpicture}

\begin{axis}[%
width=4.2in,
height=2.2in,
scale only axis,
xmin=1,
xmax=29,
xlabel = {\large utility loss/maximum distorsion:  $\max_{y \in \range{Y}}  |y-\hat{c}(y)| $},
ymin=0.5,
ymax=5,
ylabel = {\large privcy leakage: $\widetilde{\Leak}_\star(X\rightarrow Y)$},
legend style={at={(0.98,0.95)},draw=darkgray!60!black,fill=white,legend cell align=left}
]

\addplot [
color=blue,
solid,
line width=1.5pt,
mark=asterisk,
mark options={solid},
mark size=3.0pt,
]
table[row sep=crcr]{
	1 3.18078800390296\\
	2 2.97024908134218\\
	4 2.77769673768332\\
	5 2.69678188592125\\
	7 2.50678996413429\\
	8 2.17214097529461\\
	10 1.97654280542592\\
	11 1.92319923320848\\
	13 1.90790265811294\\
	14 1.84504099588056\\
	16 1.7165071265727\\
	17 1.46081152791599\\
	19 1.40247152219791\\
	20 1.41005348168399\\
	22 1.1716113780702\\
	23 1.19307644139034\\
	25 1.18813008223396\\
	26 1.18198532874354\\
	28 1.18198532874354\\
	29 0.809522758740314\\
};
\addlegendentry{\large $Y=$``blood pressure"};

\addplot [
color=red,
solid,
line width=1.5pt,
mark=diamond,
mark options={solid},
mark size=3.0pt,
]
table[row sep=crcr]{
	1 4.6903562558438\\
	2 4.07533487904045\\
	4 3.81905434738382\\
	5 3.64257645969396\\
	7 3.43162483235461\\
	8 3.347695248365\\
	10 3.22801454448293\\
	11 3.12476369915608\\
	13 3.10199150702262\\
	14 2.99302732624578\\
	16 2.94024740695002\\
	17 2.90052566422001\\
	19 2.74781361965927\\
	20 2.75698937048654\\
	22 2.59701084262027\\
	23 2.59258697941953\\
	25 2.58834100824186\\
	26 2.51985338017119\\
	28 2.47050587704492\\
	29 2.46655643214726\\
};
\addlegendentry{\large $Y=$``cholesterol"};

\end{axis}
\end{tikzpicture}%
}
	\caption{The privacy-utility tradeoff: The maximal stochastic leakage $\widetilde{\LeakS}(X\rightarrow Y)$ vs. the maximal distortion between $Y$ and $\hat{Y}$. The plots are obtained for the experiment in Section~\ref{sec:experiment} on the heart disease dataset from the UCI machine learning repository \cite{UCI2007}.}
	\label{fig:PUTLeakSProb}
\end{figure}
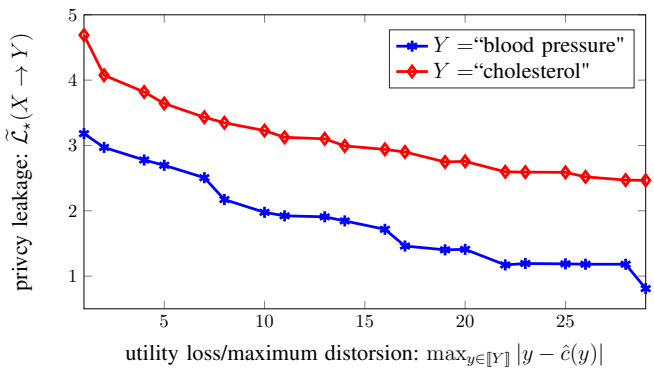

\section{Experiment}
\label{sec:experiment}

In the the UCI machine learning repository \cite{UCI2007}, the heart disease dataset was created by the Hungarian Institute of Cardiology, Budapest. It records $293$ patients' data of $76$ attributes for the purpose of identifying the presence of heart disease.
We extract three attributes in this experiment. We use the column `age' for $X$ while considering two attributes for $Y$. First, we use $Y=$``resting blood pressure (in mm Hg)'' and then switch to $Y=$``serum cholesterol (mg/dl)''.

Let $Y$ be quantized before being published.  Adopting a uniform quantizer $\hat{c}$ such that
\begin{equation}  \label{eq:Quantizer}
	\hat{c}(y)  = \delta   \Big(  \Big\lfloor \frac{y}{\delta}  \Big\rfloor  + \frac{1}{2} \Big),
\end{equation}
where $\delta>0$ refers to the \emph{step size} or \textit{resolution}. The quantized $Y$ is denoted by $\hat{Y}$ with the range $\range{\hat{Y}} = \Set{\hat{c}(y) \colon y \in \range{Y} }$.
The maximum distortion $\max_{y \in \range{Y}}  |y-\hat{c}(y)|$, capturing the loss in data utility, grows with the step size $\delta$.
We observe private information leakage from $X$ to the quantized data $\hat{Y}$.
While varying $\delta$ from $1$ to  $50$, we can compute the maximal non-stochastic brute-force guessing leakage $\Leak( X \rightarrow \hat{Y})$ and the maximin information $I_\star(X;\hat{Y})$ for each value of $\delta$ and plot them as functions of the maximum distortion in Fig.~\ref{fig:PUTLeakS}.
A tradeoff between the privacy and data utility can be seen. Both $\Leak(X \rightarrow \hat{Y})$ and $I_\star(X ; \hat{Y})$  decrease as the maximum distortion increases, i.e., the released data $\hat{Y}$ is more private as the step size of the quantization increases.
As expected from Proposition~\ref{prop:maximin}, $\Leak(X \rightarrow \hat{Y})$ is greater than $I_\star(X ; \hat{Y})$.

In Fig.~\ref{fig:PUTLeakSProb}, we also plot the privacy-utility tradeoff between the maximal stochastic leakage $\widetilde{\LeakS}(X \rightarrow \hat{Y})$ and the  maximum distortion. As seen from~\eqref{eq:RelationLeak}, $\widetilde{\LeakS}(X \rightarrow \hat{Y})$ is upper bounded by $I_\star(X ; \hat{Y})$.
Note, in this experiment, $I_\star(X ; \hat{Y}) \approx \min \Set{H_0(X), H_0(\hat{Y})}$.


%

\section{Conclusions and Future Work} \label{sec:conc}
We developed an interpretable notion of non-stochastic information leakage based on guessing in a non-stochastic framework. We considered brute-force guessing in which an adversary can potentially check all the possibilities of the private information that are compatible with the available outputs to find the actual private realization. The ratio of the worst-case number of guesses for the adversary in the presence of the output and in the absence of it captures the reduction in the adversary's guessing  complexity and is thus used as a measure of information leakage. We computed the maximal non-stochastic leakage over all sensitive attributes that could be targeted by the adversary and compared it with non-stochastic identifiability, maximin information, and stochastic maximal leakage.

One interesting finding in Example~\ref{ex:VoteMain} is that the maximal leakage is not incurred when the adversary tries to infer the exact value of $X$, but some function of $X$ that is highly related to the conditional range $\range{X|Y}$ and can be determined once $\range{Y|X}$ is fixed. An interesting question is whether this is the case in the stochastic setting and, if so, characterize the worst-case attribute, e.g., some expression in the form of \eqref{eq:WorstCaseAttri}. In stochastic leakage studies in \cite{issa2018operational,liao2018tunable}, the worst-case leakage is shown to be determined by the randomization scheme. However, the worst-case attribute has not been identified.
It is also of interest to determine whether the maximal non-stochastic brute-force guessing leakage derived in this paper and $\alpha$-leakage proposed in \cite{liao2018tunable} can be both formulated by R\'enyi measure. {The order $\alpha$ in R\'enyi entropy varies from $0$ to $\infty$, where $\alpha = 0$ refers to a non-stochastic measure. But, the $\alpha$-leakage in \cite{liao2018tunable} is only defined in $\alpha \in [1, \infty)$ based on the Arimoto mutual information. Recall that the Arimoto mutual information is defined in the whole range $\alpha \in [0,\infty)$. The question is how $\LeakS(X \rightarrow Y)$ relates to the case $(\alpha=)0$-leakage and whether the Arimoto mutual information also has a similar interpretation as in the information leakage in $\alpha \in [0,1)$. }

Another direction for future research could be to extend this definition to a dynamic framework with continual observations. In this case, we get
\begin{align*}
    X_{t+1}&=f(X_t,W_t),\\
    Y_t&=g(X_t,V_t),
\end{align*}
where $W_t$ and $V_t$ are mutually unrelated uvs. In this case, it would be interesting to understand the behaviour of $\LeakS((X_t: t\in\{1,\dots,k\}) \rightarrow (Y_t: t\in\{1,\dots,k\}))$ as a function of time and whether it can be written in recursive form. This enables us to understand private information leakage in time-varying environments.

\appendices

\section{Fundamental Properties of Information Leakage}
\label{app:axiom}

As outlined in~\cite{issa2018operational}, an operational measure of information leakage $\Leak(X \rightarrow Y)$ should hold the following basic properties:
\begin{enumerate}[R1]
	\item \emph{Cogent operational interpretation}: The leakage measure should quantify the adversary's difficulty in inferring the private/sensitive data;
	\item \emph{Minimum assumption}: Assumptions about the adversary should be minimized;
	\item Satisfying the \emph{axiomatic properties}:
	\begin{enumerate}[a)]
		\item \underline{data processing inequality}: $\Leak(X \rightarrow Y) =\min \{\Leak(X \rightarrow Z), \Leak(Y \rightarrow Z)\}$ for any $X, Y, Z$ forming a Markov chain;
		\item \underline{independence}: $\Leak(X \rightarrow Y) = 0$ if $X \perp Y$;
		\item \underline{additivity}: $\Leak(X^n \rightarrow Y^n) = n \Leak(X \rightarrow Y)$ for independently and identically distributed $(X^n,Y ^n)$.
	\end{enumerate}
	\item \emph{Accordance with intuition}: The measure should be able to identify the severity of the information leakage.
\end{enumerate}

\section{Proof of Proposition~\ref{prop:axiom}}
\label{app:prop:axiom}

\begin{IEEEproof} Proof of (a): Note that $\Leak(U \rightarrow Y)\geq 0$ for all $U$ such that $U-X-Y$ is a Markov chain; see Proposition~\ref{prop:Bounding_Leakage}.
Taking maximum of both sides of this inequality results in (a).

Proof of (b): According to Proposition~\ref{prop:Bounding_Leakage}, for unrelated $X$ and $Y$, $\Leak(U \rightarrow Y) = 0$ for all $U$ such that $U-X-Y$ is a Markov chain. Hence, $\LeakS(X \rightarrow Y) = 0$. Now, we prove the reverse. Assume that $\LeakS(X \rightarrow Y) = 0$. This implies that $\Leak(U \rightarrow Y) = 0$ for all $U$ such that $U-X-Y$ is a Markov chain. For the special case that $U = X$,
 $\Leak(U \rightarrow Y)=\Leak(X \rightarrow Y)={\max_{y\in\range{Y}} \log (|\range{X}|/ |\range{X|Y(\omega)=y}|)} = 0$ and hence we must have $|\range{X|Y(\omega)=y}| = |\range{X}|$ 
for all $y\in\range{Y}$. Noting that $\range{X|Y(\omega)=y}\subseteq\range{X}$ {and therefore} $|\range{X|Y(\omega)=y}| = |\range{X}|$ implies that $\range{X|Y(\omega)=y}= \range{X}$. Hence, $X$ and $Y$ must be unrelated.

Proof of (c): Notice that we have $\Leak(U\rightarrow Y)=\max_{y\in \range{Y}} \log
\left(|\range{U}|/|\range{U|Y(\omega)=y}|\right)\leq \log(|\range{U}|)$ because $|\range{U|Y(\omega)=y}|\geq 1$. Further, we have $|\range{U}|\leq |\range{X}|$. Hence, $\Leak(U\rightarrow Y)\leq \log(|\range{X}|)=H_0(X)$ for all $U$. Taking maximum of left hand side of this inequality over all $U$ results in (c). For $Y=X$, $\Leak(U\rightarrow Y)={ \Leak(U\rightarrow X) =}\max_{x\in \range{X}} \log
\left(|\range{U}|/|\range{U|X(\omega)=x}|\right)$. Note that $\range{U|X(\omega)=x}=\{g(x)\}$ is a singleton and, as a result, $|\range{U|X(\omega)=x}|=1$. This implies that $\Leak(U\rightarrow Y)=|\range{U}|$. Further, $|\range{U}|\leq |\range{X}|$ with equality achieved if $U=X$. Thus,
$\LeakS(X\rightarrow Y)=\sup_{U\colon U-X-Y}\Leak(U\rightarrow Y)=H_0(X).$

Proof of (d): For $U$ that holds Markov Chain $U-X-Y-Z$, we have $\Leak(U\rightarrow Y) \geq \Leak(U\rightarrow Z)$. Taking maximum of both sides of this inequality results in (d).

Proof of (e): We have $\Leak((U_i)_{i=1}^n\rightarrow (Y_i)_{i=1}^n)=\sum_{i=1}^n \Leak(U_i\rightarrow Y_i)$ if $(U_i,X_i,Y_i)$, $\forall i$, are unrelated~\cite{farokhi2019noiseless}. Note that, by definition, $(U_i,X_i,Y_i)$, $\forall i$, are unrelated if $(X_i,Y_i)$, $\forall i$, are unrelated. Taking maximum from both sides of this equality over $(U_i)_{i=1}^n$, such that $(U_i)_{i=1}^n-(X_i)_{i=1}^n-(Y_i)_{i=1}^n$ forms a Markov chain, proves (e).
\end{IEEEproof}

\section{Proof of Proposition~\ref{prop:Bounds_on_Maximal_Leakage}}
\label{app:prop:Bounds_on_Maximal_Leakage}

\begin{IEEEproof}
We start by proving that $\LeakS(X\rightarrow Y)\leq \log(|\range{X}| - \min_{y\in\range{Y}} |\range{X|Y(\omega)=y}| + 1)$. To do so, we need to prove that, $\forall y\in\range{Y}$,
\begin{align}\label{eqn:middle_result}
|\range{X}| - |\range{U}| \geq |\range{X|Y(\omega)=y}| - |\range{U|Y(\omega)=y}|.
\end{align}
This is done by \textit{reductio ad absurdum}. Assume that~\eqref{eqn:middle_result} does not hold for all $y\in\range{Y}$. Therefore, there must exists $y\in\range{Y}$ such that
\begin{align} \label{eqn:reductio_ad_absurdum}
|\range{X}| - |\range{U}| < |\range{X|Y(\omega)=y}| - |\range{U|Y(\omega)=y}|,
\end{align}
Subtracting $|\range{U}\setminus \range{U|Y(\omega)=y}|$ from both sides of~\eqref{eqn:reductio_ad_absurdum} results in
\begin{align*}
|\range{X}| - |\range{U}|-|\range{U}&\setminus \range{U|Y(\omega)=y}| \\
<& |\range{X|Y(\omega)=y}| - |\range{U|Y(\omega)=y}|\\
&-|\range{U}\setminus \range{U|Y(\omega)=y}|\\
=& |\range{X|Y(\omega)=y}| - |\range{U}|,
\end{align*}
where the equality follows from that $|\range{U}|=|\range{U|Y(\omega)=y}|+|\range{U}\setminus \range{U|Y(\omega)=y}|$ because $(\range{U}\setminus \range{U|Y(\omega)=y})\cap \range{U|Y(\omega)=y}=\emptyset$ and $(\range{U}\setminus \range{U|Y(\omega)=y})\cup \range{U|Y(\omega)=y}=\range{U}$. Therefore, it must be that
\begin{align*}
|\range{X}|-|\range{U}\setminus \range{U|Y(\omega)=y}|< |\range{X|Y(\omega)=y}|.
\end{align*}
or equivalently
\begin{align*}
|\range{X}|- |\range{X|Y(\omega)=y}|<|\range{U}\setminus \range{U|Y(\omega)=y}|.
\end{align*}
Because $(\range{X}\setminus\range{X|Y(\omega)=y})\cap \range{X|Y(\omega)=y}=\emptyset$ and ${\range{X}=}(\range{X}\setminus\range{X|Y(\omega)=y})\cup \range{X|Y(\omega)=y}$, we have $|\range{X}|=|\range{X}\setminus\range{X|Y(\omega)=y}|+|\range{X|Y(\omega)=y}|$.
Therefore, it must be that
\begin{align} \label{eqn:middle_result:1}
|\range{X}\setminus\range{X|Y(\omega)=y}|<|\range{U}\setminus \range{U|Y(\omega)=y}|.
\end{align}
On the other hand, we have
\begin{equation}
     \begin{aligned} \nonumber
         &|\range{U}\setminus \range{U|Y(\omega)=y}| \\
         & \qquad\quad = \big| \{g(x) \colon x \in \range{X} \} \setminus \{g(x) \colon x \in \range{X|Y(\omega)=y}\} \big| \\
         & \qquad\quad  \leq \big| \{g(x) \colon x \in \range{X} \setminus \range{X|Y(\omega)=y} \} \big| \\
         & \qquad\quad \leq \big| \range{X} \setminus \range{X|Y(\omega)=y} \big|,
     \end{aligned}
\end{equation}
which contradicts \eqref{eqn:middle_result:1}. Thus,~\eqref{eqn:middle_result} must be valid for all $y\in\range{Y}$.

Using~\eqref{eqn:middle_result}, we get
    \begin{equation} \nonumber
        \begin{aligned}
            \frac{|\range{U}|}{|\range{U|Y(\omega)=y}|}
            & \leq \frac{|\range{X}| - |\range{X|Y(\omega)=y}|}{|\range{U|Y(\omega)=y}|} + 1 \\
            & \leq |\range{X}| - |\range{X|Y(\omega)=y}| + 1, \; \forall y\in\range{Y},
        \end{aligned}
    \end{equation}
    where the last inequality holds because
    \begin{align}
    |\range{U|Y(\omega)=y}|
    &=\left|\bigcup_{x\in\range{X|Y(\omega)=y}}\range{U|X(\omega)=x}\right|\geq 1.\label{eqn:proof:3}
    \end{align}
    Using $y^* \in \argmin_{y\in\range{Y}} |\range{U |Y(\omega)=y}|$, we get
        \begin{align}
            \Leak(U\rightarrow Y) &= \log\left( \frac{|\range{U}|}{|\range{U|Y(\omega)=y^*}|}\right)  \nonumber\\
            & \leq \log (|\range{X}| - |\range{X|Y(\omega)=y^*}| + 1) \nonumber\\
            & =\log (|\range{X}|\hspace{-.03in} -\hspace{-.03in} \min_{y\in\range{Y}}|\range{X|Y(\omega)=y}| + 1).\label{eq:LUBaux}
        \end{align}
    Since inequality \eqref{eq:LUBaux} holds for all $U$, we have the proved upper bound.

    Now, we continue by proving the lower bound that $\LeakS(X\rightarrow Y)\geq \log(|\range{X}|-\min_{y\in\range{Y}}|\range{X|Y(\omega)=y}|+1)$. Select an arbitrary $y^*\in \argmin_{y\in\range{Y}}|\range{X|Y(\omega)=y}|$. Let us define two sets $\mathcal{X}_1:=\range{X|Y(\omega)=y^*}$ and $\mathcal{X}_2:=\range{X}\setminus\mathcal{X}_1$. Define
    $g:\range{X}\rightarrow\range{U}$ with $\range{U}=\mathcal{X}_2\cup\{u^*\}$ as
    \begin{align} \label{eq:WorstCaseAttri}
        g(x)
        =
        \begin{cases}
            u^*, & x\in\mathcal{X}_1,\\
            x, & x\in\mathcal{X}_2.
        \end{cases}
    \end{align}
    Note that, by construction, $|\range{U|Y(\omega)=y^*}|=|\{u^*\}|=1$ and $|\range{U|Y(\omega)=y}|=|g(\range{X|Y(\omega)=y})|\geq 1$ for all $y\in\range{Y}\setminus\{y^*\}$. Hence, $\min_{y\in\range{Y}} |\range{U|Y(\omega)=y}|=1$. Therefore,
    \begin{align*}
        \LeakS(X\rightarrow Y)
        &\geq \Leak(U\rightarrow Y)\\
        &=\log\left( \frac{|\range{U}|}{ \displaystyle\min_{y \in \range{Y}} |\range{U|Y(\omega)=y}|}\right) \\
        &=\log(|\range{U}|)\\
        &=\log(|\range{X}\setminus \range{X|Y(\omega)=y^*}|+1)\\
        &=\log(|\range{X}|-|\range{X|Y(\omega)=y^*}|+1)\\
        &=\log(|\range{X}|-\min_{y\in\range{Y}}|\range{X|Y(\omega)=y}|+1).
    \end{align*}
    This concludes the proof.
\end{IEEEproof}

\section{$\LeakS(X \rightarrow Y)$ and $\Leak(X \rightarrow Y )$}
\label{app:eq:LeakLeakS}

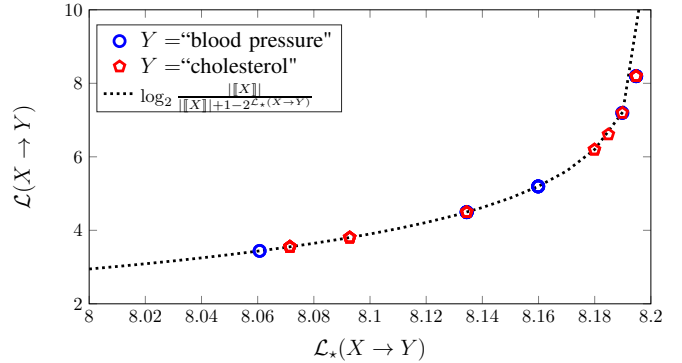
\begin{figure}[t]
	\centering
	\scalebox{0.7}{
%
%
\begin{tikzpicture}

\begin{axis}[%
width=4.2in,
height=2.2in,
scale only axis,
xmin=8,
xmax=8.2,
xlabel = {\large $\LeakS(X \rightarrow Y)$},
ymin=2,
ymax=10,
ylabel = {\large $\Leak(X \rightarrow Y)$},
legend style={at={(0.45,0.95)},draw=darkgray!60!black,fill=white,legend cell align=left}
]

\addplot [
color=blue,
line width=1.5pt,
only marks,
mark=o,
mark options={solid},
mark size=3.0pt,
]
table[row sep=crcr]{
8.19475685442225 8.19475685442225\\
8.19475685442225 8.19475685442225\\
8.19475685442225 8.19475685442225\\
8.19475685442225 8.19475685442225\\
8.19475685442225 8.19475685442225\\
8.18982455888002 7.19475685442225\\
8.18982455888002 7.19475685442225\\
8.15987133677839 5.19475685442225\\
8.15987133677839 5.19475685442225\\
8.15987133677839 5.19475685442225\\
8.15987133677839 5.19475685442225\\
8.15987133677839 5.19475685442225\\
8.13442632022093 4.49431713628116\\
8.13442632022093 4.49431713628116\\
8.13442632022093 4.49431713628116\\
8.13442632022093 4.49431713628116\\
8.13442632022093 4.49431713628116\\
8.13442632022093 4.49431713628116\\
8.06069593168755 3.43986935225878\\
8.06069593168755 3.43986935225878\\
};
\addlegendentry{\large $Y=$``blood pressure"};

\addplot [
color=red,
line width=1.5pt,
only marks,
mark=pentagon,
mark options={solid},
mark size=3.0pt,
]
table[row sep=crcr]{
8.19475685442225 8.19475685442225\\
8.19475685442225 8.19475685442225\\
8.19475685442225 8.19475685442225\\
8.18982455888002 7.19475685442225\\
8.18487534290828 6.60979435370109\\
8.17990909001493 6.19475685442225\\
8.17990909001493 6.19475685442225\\
8.13442632022093 4.49431713628116\\
8.13442632022093 4.49431713628116\\
8.13442632022093 4.49431713628116\\
8.09275714091985 3.80243943164349\\
8.09275714091985 3.80243943164349\\
8.09275714091985 3.80243943164349\\
8.09275714091985 3.80243943164349\\
8.07146236255662 3.55090066464752\\
8.07146236255662 3.55090066464752\\
8.07146236255662 3.55090066464752\\
8.07146236255662 3.55090066464752\\
8.07146236255662 3.55090066464752\\
8.07146236255662 3.55090066464752\\
};
\addlegendentry{\large $Y=$``cholesterol"};

\addplot [
color=black,
dotted,
line width=1.5pt,
]
table[row sep=crcr]{
8 2.94682934097866\\
8.01 3.0160669413999\\
8.02 3.08931455887246\\
8.03 3.16703235836113\\
8.04 3.2497644326819\\
8.05 3.33816054090263\\
8.06 3.43300537519491\\
8.07 3.53525871921854\\
8.08 3.64611175585586\\
8.09 3.7670680044478\\
8.1 3.90006306542347\\
8.11 4.04764789210537\\
8.12 4.21328088562856\\
8.13 4.4018169333335\\
8.14 4.62037793759018\\
8.15 4.88002913344999\\
8.16 5.1993641185599\\
8.17 5.61339887718363\\
8.18 6.20136313845212\\
8.19 7.22060295750093\\
8.2 12.0992211635436\\
};
\addlegendentry{$\log_2 \frac{|\range{X}|}{|\range{X}| + 1 - 2^{\LeakS(X \rightarrow Y)}}$};
\end{axis}
\end{tikzpicture}%
}
	\caption{The identifiability $\Leak(X \rightarrow Y)$ on $X$ as a function of the maximal non-stochastic brute-force guessing leakage $\LeakS(X \rightarrow Y)$ for the heart disease dataset  in the UCI machine learning repository \cite{UCI2007}: $Y =$``blood pressure" and $Y = $``cholesterol". }
	\label{fig:Identifiablity}
\end{figure}

While Corollary~\ref{cor:ident} shows that the maximal non-stochastic guessing leakage $\LeakS(X \rightarrow Y)$ is upper bounded by the privacy budget $\epsilon$ for the non-stochastic identifiability $\Leak(X \rightarrow Y)$, we can show the more general result that $\Leak(X \rightarrow Y)$ is monotonic in $\LeakS(X \rightarrow Y)$; see \eqref{eq:LeakLeakS}.
For the Hungarian heart disease dataset used in Section~\ref{sec:experiment}, we can plot $\Leak(X \rightarrow Y)$ vs $\LeakS(X \rightarrow Y)$ for different quantization levels as in Fig.~\ref{fig:Identifiablity}. The logarithm here is in base of $2$. The plot aligns with \eqref{eq:LeakLeakS}, i.e.,  $\Leak(X \rightarrow Y) = \log_2 ({|\range{X}|}/{|\range{X}| + 1 - 2^{\LeakS(X \rightarrow Y)}})$. We can see that $\Leak(X \rightarrow Y)$ increases with $\LeakS(X \rightarrow Y)$.

\bibliography{citation}
\bibliographystyle{IEEEtran}

\end{document}